\documentstyle[aps,12pt,epsf,psfig,preprint]{revtex}
\begin{document}

\draft

\title{A unique large thermal source of real and virtual photons in
the reactions Pb(158 AGeV) + Pb, Au
}

\author{
{\sc
K. Gallmeister$^a$, B. K\"ampfer$^a$, O.P. Pavlenko$^{a,b}$}
}

\address{
$^a$Forschungszentrum Rossendorf, PF 510119, 01314 Dresden, Germany\\
$^b$Institute for Theoretical Physics, 252143 Kiev - 143, Ukraine
}

\maketitle

\begin{abstract}
The data of direct single-photon measurements of
the WA98 collaboration in the reaction Pb(158 AGeV) + Pb
are analyzed within a thermal model with a minimum number of
parameters adjusted to the dilepton data obtained by the CERES and
NA50 collaborations in the reactions Pb(158 AGeV) + Au, Pb.
The agreement of our model with the WA98 data points to a unique 
large thermal source emitting electromagnetic radiation observable
in both the real and virtual photon channels.
\end{abstract}

\pacs{\\ %\vskip 1cm
{\it Key Words:\/}
heavy-ion collisions, photons, deconfinement\\
{\it PACS:\/}
25.75.+r, 12.38.Mh, 24.85.+p}

\section{Introduction}

Direct real and virtual photons are now widely considered as very
important penetrating probes of the early, hot and dense stage of
relativistic heavy-ion collisions
(cf. \cite{Rapp_Wambach} for a recent survey). 
Efforts \cite{Rapp_Shuryak,PL_2000,Hirschegg_2000} have been devoted
recently to understand the dielectron spectra measured by the
CERES collaboration in the reaction Pb(158 AGeV) + Au in the
low invariant mass region \cite{CERES} and the dimuon spectra
measured by the NA50 collaboration in the reaction Pb(158 AGeV) + Pb
in the intermediate mass region \cite{NA50}.
Both experiments point to a significant excess of dileptons above
the conventional sources such as electromagnetic decays of light
hadrons (resulting in the hadronic cocktail), and correlated
semileptonic decays of open charm mesons, and the Drell-Yan process.
It is commonly believed that, in particular, the CERES data,
obtained with lead and sulfur projectiles, could
indicate a drastic change of the properties of hadrons, mainly the
$\rho$ meson, in a hot and dense surrounding hadronic medium
\cite{Rapp_Wambach}.
For instance, transport simulations \cite{Brown_Li,Cassing}
including in-medium modifications of the light vector meson masses and
widths are found to account for the excess of dielectrons. At the same time 
''traditional'' hydrodynamical calculations \cite{Huovinen_Prakash},
even when supplemented by the in-medium effects, give a dilepton yield
which appears to be below the experimental data in the most important
invariant mass region, 0.4 GeV $< M <$ 0.6 GeV. Only if one employs 
chemical off-equilibrium effects, e.g., parameterized by a finite
pion chemical potential \cite{our_old}, the agreement of the
hydrodynamical model calculations with data can be improved
\cite{Rapp_Shuryak}.

In a recent analysis \cite{PL_2000,Hirschegg_2000} we proposed a 
model with a minimum set of parameters, which describes at the same time
the CERES and NA50 dilepton data by a thermal source. In spite of its
schematic character, the model is based on the nontrivial observation,
made recently as a result of rather involved evaluations of the
properties of hadrons near the chiral symmetry restoration, that the
dilepton emission rate from a hadron gas at given temperature
is fairly well described
by the quark--anti-quark annihilation rate at the same temperature
in a wide range of invariant masses extending to small values
\cite{Rapp_Wambach}. 
This is called a ''duality'' of the hadronic and quark-gluon degrees
of freedom in describing the emissivity of strongly interacting
matter. Such a simplification makes it possible to parameterize the
thermal source by a Boltzmann like exponential function with effective
temperature $T_{\rm eff}$ and a normalization factor $N_{\rm eff}$
which reflects the space-time volume occupied by the thermal source
\cite{Hirschegg_2000}. 
The most remarkable result of such a parameterization is that the
dilepton spectra in both the low-mass region, as measured by the CERES 
collaboration, and the intermediate-mass region, as accessible to
the NA50 collaboration, can be well described by the two unique
parameters $T_{\rm eff}$ and $N_{\rm eff}$. Also the transverse
momentum spectra are well described \cite{PL_2000,Hirschegg_2000}.
This clearly indicates a common thermal source seen in different phase
space regions and different dilepton channels.

In the present paper we employ the above mentioned model for the
thermal source to analyze the direct real photons in the
reaction Pb(158 AGeV) + Pb. In line with the hadron-quark ''duality''
we assume that deconfined and hadron matter shine equally bright in
a temperature region centered around the critical temperature
of chiral symmetry restoration. 

An important information, which can be extracted from the photon
spectra, is related to the transverse collective flow of matter.
Unfortunately, the corresponding transverse momentum ($Q_\perp$)
spectra of dileptons do not give so far such an opportunity,
since in the large $Q_\perp$ region, where the flow effect is strongest,
the total yield is dominated by the background contributions.

The primary aim of the present note is to show that a thermal source,
with parameters adjusted to the CERES and NA50 dilepton data, 
results in a photon spectrum which, when
including the background, is in agreement
with the data of the WA98 collaboration \cite{WA98} for the
reaction Pb(158 AGeV) + Pb.
 
\section{Model}    
 
For the emission of photons from a thermal source we employ the rate
obtained in \cite{Kapusta} for the free quark-gluon plasma at
temperature $T$
\begin{equation}
E \frac{dN}{d^4 x \, d^3 p} =
\frac 59 \frac{\alpha \alpha_s}{2 \pi^2} T^2 
\exp \left\{ - \frac{p \cdot u}{T} \right\}
\log \left[ 1 + \frac{\kappa (p \cdot u)}{\alpha_s T} \right],
\label{eq.1}
\end{equation}
where $p^\mu = (E, \vec p) = (p_\perp \cosh y, p_\perp \sinh y, \vec
p_\perp)$
denotes the four-momentum of the photon with transverse momentum
$p_\perp$
and rapidity $y$; $u^\mu$ is the four-velocity of the medium;
the constant $\kappa$ is according to \cite{Kapusta}
$\kappa = 2.912/(4 \pi)$. We use units with $\hbar = c =1$.
The rate (\ref{eq.1}) takes into account the lowest order QCD
processes, i.e. the $q \bar q$ annihilation and the Compton like process.
For the strong coupling strength we take $\alpha_s = 0.3$.
As shown in \cite{Kapusta}, at $T = 200$ MeV this rate is very close
to that of hadron matter when including a quite complete list of
reactions of light mesons and the decays of vector mesons. This gives
an additional hint for extending the mentioned hadron-quark
''duality'' to the real photon sector, at least for temperatures in
the region of the chiral symmetry restoration. 
(At much higher temperature, the rate in \cite{Aurenche} is probably
more appropriate; cf.\ also \cite{Srivastava}.)

Since the thermal photon spectra are thought to be sensitive to the
medium's flow we consider here the simplest case of a spherically
symmetric expansion with velocity profile 
$v(r) = v_0 r / R(t)$, where $r$ stands for the radial coordinate and
$R(t)$ is the radius of the source, so that for a constant velocity
parameter $v_0$ the size $R(t)$ increases linearly with time $t$.
In line with our model of the thermal dilepton source 
\cite{PL_2000,Hirschegg_2000}
we replace the temperature by an effective average temperature 
$T_{\rm eff}$, being constant.
Within such an approximation one can factorize the space-time
volume $V_4$ and the radiation emissivity. Performing the
space-time integration in eq.~(\ref{eq.1}) one gets the photon
spectrum as
\begin{eqnarray}
E \frac{dN}{d^3 p} & = &
V_4 \,  F (E,T_{\rm eff}, v_0), 
\label{eq.2} \\
F & = & 
\frac{3}{4 \pi} \int d^3 \vec s 
\frac{dN}{d^4 x \, (d^3 \vec p /E)},
\label{eq.3}
\end{eqnarray} 
where $\vec s = \vec r / R(t)$ and 
$V_4 = \frac{4 \pi}{3} \int dt \, R(t)^3$.
Below we use the uniquely fixed parameters
$T_{\rm eff} = 170$ MeV and $V_4 = N_{\rm eff} = 3.3 \times 10^4$ fm${}^4$,
which have been previously adjusted \cite{PL_2000,Hirschegg_2000}
to the CERES and NA50 dilepton data.

\section{Analysis of photon spectra}

We focus on the direct photon production assuming that the
background, related to decays of secondary hadrons, is removed from the 
corresponding data \cite{WA98}. To describe the spectra one also needs
the contribution from hard initial processes, like the Drell-Yan
process in dilepton production, 
which is expected to dominate in the high-$p_\perp$ region.
The hard photon yield is generated in our study by the event generator
PYTHIA \cite{PYTHIA} with structure functions MRS D-', default
cut-off parameter $\hat p_{\perp min} = 1$ GeV for photons, and intrinsic
transverse parton momentum spread
$\sqrt{\langle k_\perp^2 \rangle} =$ 0.8 GeV, adjusted to the
transverse momentum spectra of dileptons in the Drell-Yan region
in \cite{PL_2000}; otherwise the default switches are used.
Higher order calculations are presented in \cite{Wong}, but
we try to absorb them into an appropriate K factor.

To check the reliability of estimates of the hard component,
we compare in fig.~1 the PYTHIA results with the photon data obtained in 
pp collisions by the E704 collaboration at $\sqrt{s} =$ 19.4 GeV \cite{E704}
(left part) and by the UA6 collaboration at $\sqrt{s} =$ 24.3 GeV \cite{UA6}
(right part). To reproduce the data one has to include the K factors
of 3.2 (1.8) for $\sqrt{s} =$ 19.4 GeV ( $\sqrt{s} =$ 24.3 GeV).
As seen in fig.~1, PYTHIA describes then well the data, i.e. the
slopes of the spectra, which are in
the high-$p_\perp$ region, $p_\perp > 2.5$ GeV. This conclusion is
important for the subsequent analysis of the photon data in heavy-ion
collisions, where the absolute normalization of the hard production
(i.e., the number of nucleon-nucleon collisions) is 
{\it a priori} unknown. 

The comparison of our thermal model yield plus the hard yield from
PYTHIA with the WA98 data \cite{WA98} is displayed in
fig.~2 for the case of neglecting the flow, i.e., $v_0 = 0$, and
in fig.~3 for the flow parameter $v_0 = 0.3$.
(Similar to the PYTHIA calculations we simulate the thermal yield 
eqs.~(\ref{eq.2}, \ref{eq.3}) by a
Monte Carlo procedure and apply the detector acceptance.)
As seen in fig.~2, the hard photon production process dominates in the
high-$p_\perp$ region, where we have adjusted the PYTHIA calculation
to data.
At $p_\perp < 2.5$ GeV one can observe the increase
of the experimental yield above the hard yield. 
In this region secondary processes become important.
One has to stress, however, that in 
the region of not too large values of $p_\perp$ any hard photon
production calculation based on perturbative QCD is not longer
reliable. In particular, in the region of smaller $p_\perp$ the
PYTHIA results become sensitive to the cut-off parameter 
$\hat p_{\perp min}$.

As can be expected from eq.~(\ref{eq.1}), the contribution from the
thermal source strongly depends on the transverse flow.
The moderate value of $v_0 = 0.3$, used in fig.~3, is consistent
with the analysis of transverse hadron flow at kinetic decoupling
\cite{bk}. For such a flow parameter one finds good
agreement with the data \cite{WA98} when adding the thermal yield and
the hard contribution (see fig.~3).

To make more firm conclusions on the role of thermal photons in 
Pb + Pb collisions at CERN-SPS energies, one needs a more reliable
procedure to fix the hard background. Similarly to dileptons, an
accurate adjustment of the hard rate at the high-$p_\perp$ tail
requires an improvement of the data statistics. 
Notice that our present up-scale factor from PYTHIA simulations of
pp collisions to heavy-ion data can be considered as an upper bound
for the hard radiation. Nevertheless, a remarkable space is left for 
the secondary radiation.

An additional
insight can be gained by an analysis of non-central collisions
which should allow a link to pA collisions, where also data are
at hand \cite{pA}. This will be subject of a separate future study.
%when experimental data from non-central collisions become available.
%along with dilepton data from S(200 AGeV) + Au, Pb, W collisions.

\section{Summary and discussion}

In summary we analyze the direct photon production in the
reaction Pb(158 AGeV) + Pb by using a model with a minimum parameter
set, i.e., the effective temperature and the space-time volume
of the thermal source, both ones adjusted to dilepton data in similar
central reactions. The model employs the hadron-quark ''duality'' for the
rate of electromagnetic radiation off matter. We have shown that our
model, supplemented by the hard photon yield as described by PYTHIA,
is in good agreement with the WA98 data. This result
supports the assumption of an extended and long-living source of the 
electromagnetic radiation, which can be seen in both the real and
virtual photon channels. The comparatively large value of the
space-time factor $V_4$ in our model can be related to
previous expectations on a long-living fireball \cite{Shuryak}
or a phase space overpopulation of hadrons
\cite{Rapp_Shuryak,our_old}.

It should be emphasized that our effective temperature parameter
$T_{\rm eff} = 170$ MeV is in perfect agreement with the temperature
parameter needed to describe hadron species ratios
\cite{Cleymans}. Since $T_{\rm eff}$ is to be considered as
average of the temperature, one concludes that the electromagnetic
probes indeed point to temperatures above the expected deconfinement
temperature. This corroborates the expectation of exciting deconfined
matter in central heavy-ion collisions already at CERN-SPS energies.

The photon spectra are shown to be useful in extracting information
on transverse flow. For firm conclusions, however, the hard photon
production processes must be reliably accessible.
Our approach can be contrasted with attempts to interpret the data
either without any hard contribution \cite{Srivastava,Alam} or by the hard yield
alone by tuning parameters.  
Transport approaches \cite{transport} should smoothly interpolate
between these extreme cases.

We expect that the future analysis of the starting experiments at the
relativistic heavy-ion collider RHIC at Brookhaven National Laboratory
should deliver a higher value of $T_{\rm eff}$ since the
estimated maximum temperatures will be significantly larger. 

\subsection*{Acknowledgments}

Useful discussions with H.W. Barz and
G. Zinovjev
are gratefully acknowledged, in particular
Th. Peitzmann for valuable explanations of the WA98 data. 
O.P.P. thanks for the warm hospitality in the nuclear theory group
in the Research Center Rossendorf. 
The work is supported by the grants BMBF 06DR829/1,
STCU 015 and WTZ UKR-008-98.

\begin{figure}
~\\[-.1cm]
\centerline{{\psfig{file=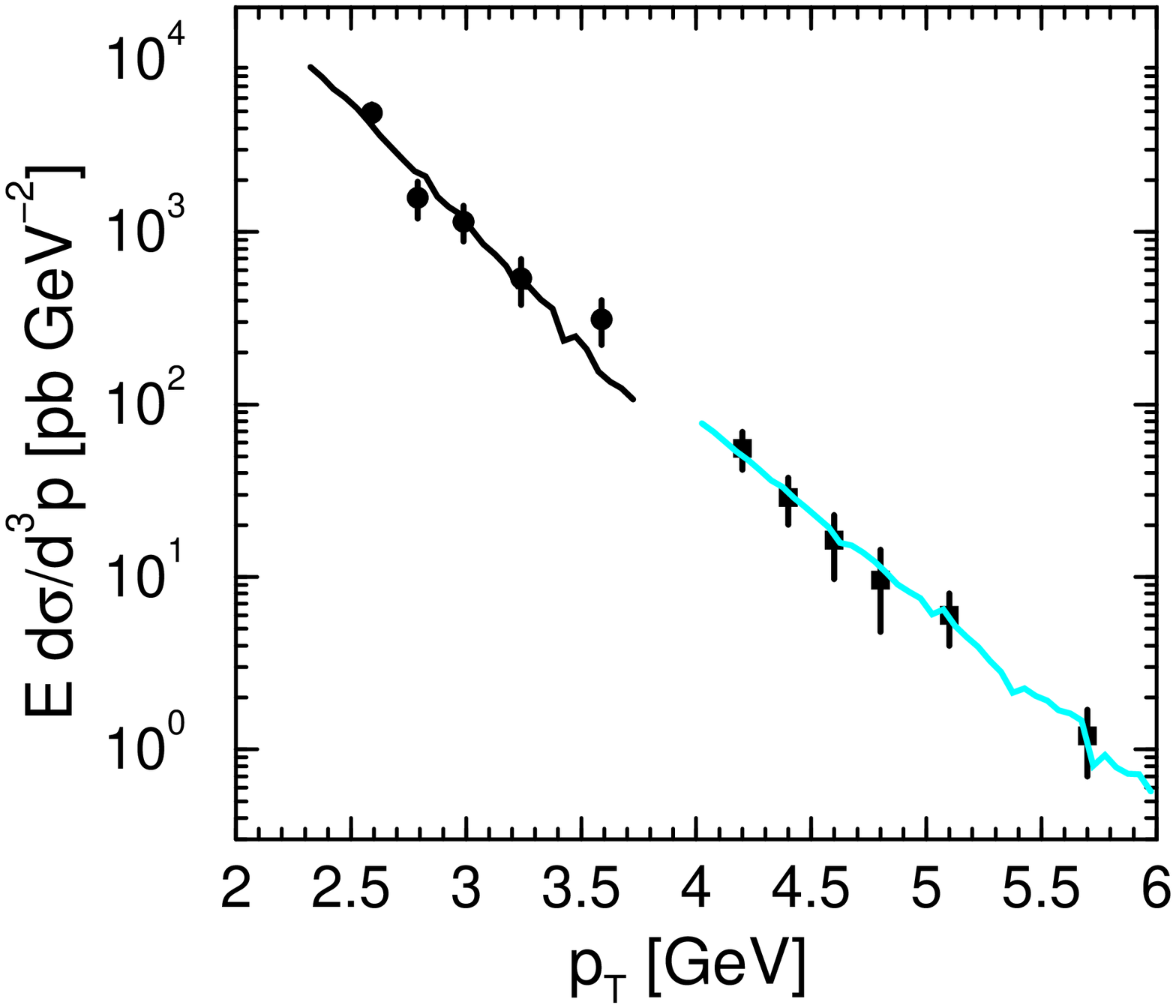,width=9cm,angle=-90}}}
\caption{
Comparison of PYTHIA results with direct photon data in pp collisions.
Left part: $\sqrt{s}$ = 19.4 GeV and E704 data (full dots) 
\protect\cite{E704},
right part: $\sqrt{s}$ = 24.3 GeV and UA6 data (full squares) 
\protect\cite{UA6}.
}
\end{figure}

\begin{figure}
~\\[-.1cm]
\centerline{{\psfig{file=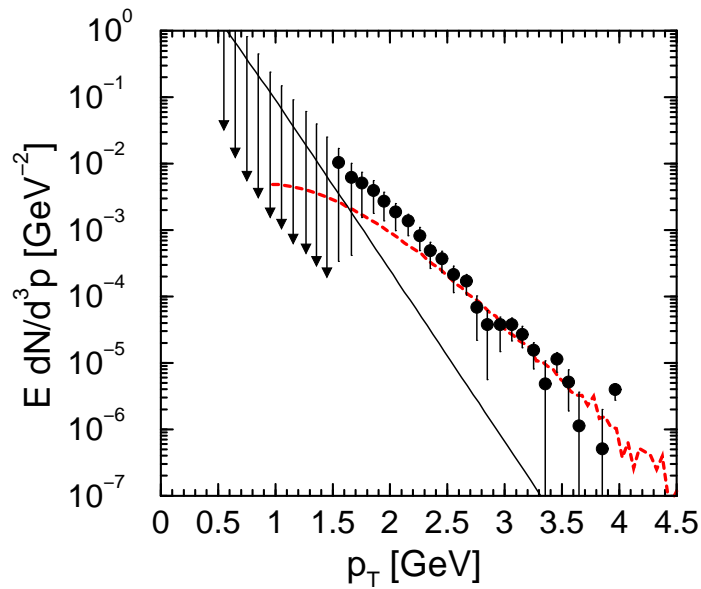,width=9cm,angle=-90}}}
\caption{
Comparison of the thermal photon yield (thin curve, no transverse
expansion) and the hard
background (dashed curve) with the direct photon data
\protect\cite{WA98} for the reaction Pb(158 AGeV) + Pb. 
}
\end{figure}

\begin{figure}
~\\[-.1cm]
\centerline{{\psfig{file=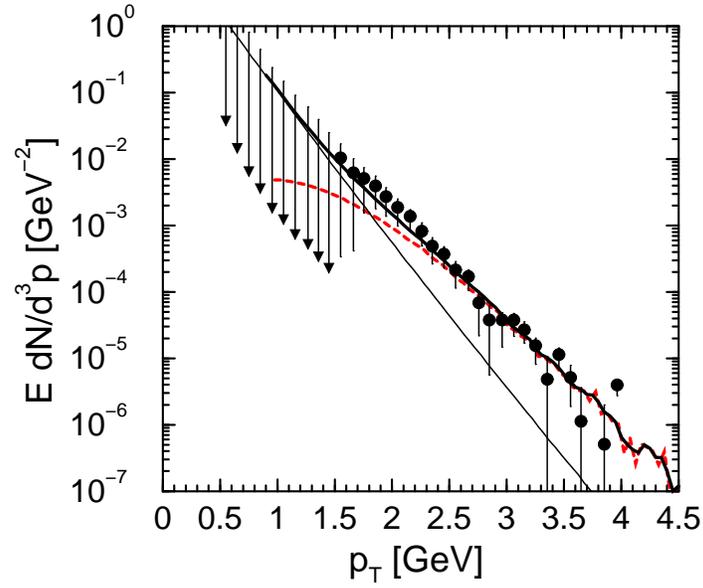,width=9cm,angle=-90}}}
\caption{
As in fig.~2, however with transverse flow ($v_0 = 0.3$).
The solid line depicts the sum of the thermal and hard yields.
}
\end{figure}
\end{document}